\documentclass[12pt]{article}
\newcommand{\nc}{\newcommand}
\nc{\renc}{\renewcommand}

\usepackage{graphicx}

\usepackage{graphicx}

\nc{\half}{{\textstyle{1\over2}}}
\nc{\etal}{\mbox{\it et al. }}
\nc{\ie}{{\it i.e.}}
\nc{\eg}{{\it e.g.}}

\renc{\thefootnote}{\arabic{footnote}}
\nc{\capt}[1]{{\bf Figure.} {\small\sl #1}}

\nc{\eqs}[2]{\mbox{Eqs.~(\ref{#1},\,\ref{#2})}}
\nc{\eq}[1]{\mbox{Eq.~(\ref{#1})}}

\nc{\figs}[2]{\mbox{Figs.~(\ref{#1},\,\ref{#2})}}
\nc{\fig}[1]{\mbox{Fig~.(\ref{#1})}}

\nc{\tag}[1]{\label{#1} \marginpar{{\footnotesize #1}}}
\nc{\mtag}[1]{\label{#1} \mbox{\marginpar{{\footnotesize #1}}}}
\renc{\baselinestretch}{1.5}
\newlength{\overeqskip}
\newlength{\undereqskip}
\nc{\be}[1]{\begin{equation} \mbox{$\label{#1}$}}
\nc{\bea}[1]{\begin{eqnarray} \mbox{$\label{#1}$}}
\nc{\Section}[2]{\section{#2}\label{#1}}
\nc{\Bibitem}[1]{\bibitem{#1}}
\nc{\Label}[1]{\label{#1}}

\nc{\eea}{\vspace{\undereqskip}\end{eqnarray}}
\nc{\ee}{\vspace{\undereqskip}\end{equation}}
\nc{\bdm}{\begin{displaymath}}
\nc{\edm}{\end{displaymath}}
\nc{\dpsty}{\displaystyle}
\nc{\bc}{\begin{center}}
\nc{\ec}{\end{center}}
\nc{\ba}{\begin{array}}
\nc{\ea}{\end{array}}
\nc{\bab}{\begin{abstract}}
\nc{\eab}{\end{abstract}}
\nc{\btab}{\begin{tabular}}
\nc{\etab}{\end{tabular}}
\nc{\bit}{\begin{itemize}}
\nc{\eit}{\end{itemize}}
\nc{\ben}{\begin{enumerate}}
\nc{\een}{\end{enumerate}}
\nc{\bfig}{\begin{figure}}
\nc{\efig}{\end{figure}}
\nc{\arreq}{&\!=\!&}
\nc{\arrmi}{&\!-\!&}
\nc{\arrpl}{&\!+\!&}
\nc{\arrap}{&\!\!\!\approx\!\!\!&}
\nc{\non}{\nonumber\\*}
\nc{\align}{\!\!\!\!\!\!\!\!&&}

\def\lsim{\; \raise0.3ex\hbox{$<$\kern-0.75em
      \raise-1.1ex\hbox{$\sim$}}\; }
\def\gsim{\; \raise0.3ex\hbox{$>$\kern-0.75em
      \raise-1.1ex\hbox{$\sim$}}\; }
\nc{\DOT}{\hspace{-0.08in}{\bf .}\hspace{0.1in}}
\nc{\Laada}{\hbox {$\sqcap$ \kern -1em $\sqcup$}}
\nc\loota{{\scriptstyle\sqcap\kern-0.55em\hbox{$\scriptstyle\sqcup$}}}
\nc\Loota{{\sqcap\kern-0.65em\hbox{$\sqcup$}}}
\nc\laada{\Loota}
\nc{\qed}{\hskip 3em \hbox{\BOX} \vskip 2ex}

\nc{\real}{{\rm I \! R}}
\nc{\Z}{{\sf Z \!\!\! Z}}
\nc{\complex}{{\rm C\!\!\! {\sf I}\,\,}}
\def\bigid{\leavevmode\hbox{\small1\kern-3.8pt\normalsize1}}
\def\id{\leavevmode\hbox{\small1\kern-3.3pt\normalsize1}}
\nc{\slask}{\!\!\!/}
\nc{\bis}{{\prime\prime}}
\nc{\pa}{\partial}
\nc{\na}{\nabla}
\nc{\ra}{\rangle}
\nc{\la}{\langle}
\nc{\goto}{\rightarrow}
\nc{\swap}{\leftrightarrow}

\nc{\EE}[1]{ \mbox{$\cdot10^{#1}$} }
\nc{\abs}[1]{\left|#1\right|}
\nc{\at}[2]{\left.#1\right|_{#2}}
\nc{\norm}[1]{\|#1\|}
\nc{\abscut}[2]{\Abs{#1}_{\scriptscriptstyle#2}}
\nc{\vek}[1]{{\rm\bf #1}}
\nc{\integral}[2]{\int\limits_{#1}^{#2}}
\nc{\inv}[1]{\frac{1}{#1}}
\nc{\dd}[2]{{{\partial #1}\over{\partial #2}}}
\nc{\ddd}[2]{{{{\partial}^2 #1}\over{\partial {#2}^2}}}
\nc{\dddd}[3]{{{{\partial}^2 #1}\over
        {\partial #2 \partial #3}}}
\nc{\dder}[2]{{{d #1}\over{d #2}}}
\nc{\ddder}[2]{{{d^2 #1}\over{d {#2}^2}}}
\nc{\dddder}[3]{{d^2 #1}\over
        {d #2 d #3}}
\nc{\dx}[1]{d\,^{#1}x}
\nc{\dy}[1]{d\,^{#1}y}
\nc{\dz}[1]{d\,^{#1}z}
\nc{\dl}[1]{\frac{d\,^{#1}l}{(2\pi)^{#1}}}
\nc{\dk}[1]{\frac{d\,^{#1}k}{(2\pi)^{#1}}}
\nc{\dq}[1]{\frac{d\,^{#1}q}{(2\pi)^{#1}}}

\nc{\cc}{\mbox{$c.c.$ }}
\nc{\hc}{\mbox{$h.c.$ }}
\nc{\cf}{cf.\ }
\nc{\erfc}{{\rm erfc}}
\nc{\Tr}{{\rm Tr\,}}
\nc{\tr}{{\rm tr\,}}
\nc{\pol}{{\rm pol}}
\nc{\sign}{{\rm sign}}
\nc{\bfT}{{\bf T }}
\def\eV{{\rm\ eV}}
\def\GeV{{\rm\ GeV}}
\def\MeV{{\rm\ MeV}}

\def\TeV{{\rm\ TeV}}

\nc{\cA}{{\cal A}}
\nc{\cB}{{\cal B}}
\nc{\cD}{{\cal D}}
\nc{\cE}{{\cal E}}
\nc{\cG}{{\cal G}}
\nc{\cH}{{\cal H}}
\nc{\cL}{{\cal L}}
\nc{\cO}{{\cal O}}
\nc{\cT}{{\cal T}}
\nc{\cN}{{\cal N}}
\nc{\rvac}[1]{|{\cal O}#1\rangle}
\nc{\lvac}[1]{\langle{\cal O}#1|}
\nc{\rvacb}[1]{|{\cal O}_\beta #1\rangle}
\nc{\lvacb}[1]{\langle{\cal O}_\beta #1 |}
\nc{\bb}{\bar{\beta}}
\nc{\bt}{\tilde{\beta}}
\nc{\ctH}{\tilde{\cal H}}
\nc{\chH}{\hat{\cal H}}
\nc{\1}{\aa}
\nc{\2}{\"{a}}
\nc{\3}{\"{o}}
\nc{\4}{\AA}
\nc{\5}{\"{A}}
\nc{\6}{\"{O}}
\nc{\al}{\alpha}
\nc{\g}{\gamma}
\nc{\Del}{\Delta}
\nc{\e}{\epsilon}
\nc{\eps}{\epsilon}
\nc{\lam}{\lambda}
\nc{\om}{\omega}
\nc{\Om}{\Omega}
\nc{\ve}{\varepsilon}
\nc{\mn}{{\mu\nu}}
\nc{\kap}{\kappa}
\nc{\vp}{\varphi}

\nc{\advp}[3]{{\it  Adv.\ in\ Phys.\ }{{\bf #1} {(#2)} {#3}}}
\nc{\annp}[3]{{\it  Ann.\ Phys.\ (N.Y.)\ }{{\bf #1} {(#2)} {#3}}}
\nc{\apl}[3]{{\it  Appl. Phys. Lett. }{{\bf #1} {(#2)} {#3}}}
\nc{\apj}[3]{{\it  Ap.\ J.\ }{{\bf #1} {(#2)} {#3}}}
\nc{\apjl}[3]{{\it  Ap.\ J.\ Lett.\ }{{\bf #1} {(#2)} {#3}}}
\nc{\app}[3]{{\it Astropart.\ Phys.\ }{{\bf #1} {(#2)} {#3}}}
\nc{\cmp}[3]{{\it  Comm.\ Math.\ Phys.\ }{{ \bf #1} {(#2)} {#3}}}
\nc{\cqg}[3]{{\it  Class.\ Quant.\ Grav.\ }{{\bf #1} {(#2)} {#3}}}
\nc{\epl}[3]{{\it  Europhys.\ Lett.\ }{{\bf #1} {(#2)} {#3}}}
\nc{\ijmp}[3]{{\it Int.\ J.\ Mod.\ Phys.\ }{{\bf #1} {(#2)} {#3}}}
\nc{\ijtp}[3]{{\it Int.\ J.\ Theor.\ Phys.\ }{{\bf #1} {(#2)} {#3}}}
\nc{\jmp}[3]{{\it  J.\ Math.\ Phys.\ }{{ \bf #1} {(#2)} {#3}}}
\nc{\jpa}[3]{{\it  J.\ Phys.\ A\ }{{\bf #1} {(#2)} {#3}}}
\nc{\jpc}[3]{{\it  J.\ Phys.\ C\ }{{\bf #1} {(#2)} {#3}}}
\nc{\jap}[3]{{\it J.\ Appl.\ Phys.\ }{{\bf #1} {(#2)} {#3}}}
\nc{\jpsj}[3]{{\it J.\ Phys.\ Soc.\ Japan\ }{{\bf #1} {(#2)} {#3}}}
\nc{\lmp}[3]{{\it Lett.\ Math.\ Phys.\ }{{\bf #1} {(#2)} {#3}}}
\nc{\mpl}[3]{{\it  Mod.\ Phys.\ Lett.\ }{{\bf #1} {(#2)} {#3}}}
\nc{\ncim}[3]{{\it  Nuov.\ Cim.\ }{{\bf #1} {(#2)} {#3}}}
\nc{\np}[3]{{\it  Nucl.\ Phys.\ }{{\bf #1} {(#2)} {#3}}}
\nc{\npps}[3]{{\it  Nucl.\ Phys.\ Proc.\ Suppl.\ }{{\bf #1} {(#2)} {#3}}}
\nc{\pr}[3]{{\it Phys.\ Rev.\ }{{\bf #1} {(#2)} {#3}}}
\nc{\pra}[3]{{\it  Phys.\ Rev.\ A\ }{{\bf #1} {(#2)} {#3}}}
\nc{\prb}[3]{{\it  Phys.\ Rev.\ B\ }{{{\bf #1} {(#2)} {#3}}}}
\nc{\prc}[3]{{\it  Phys.\ Rev.\ C\ }{{\bf #1} {(#2)} {#3}}}
\nc{\prd}[3]{{\it  Phys.\ Rev.\ D\ }{{\bf #1} {(#2)} {#3}}}
\nc{\prl}[3]{{\it Phys.\ Rev.\ Lett.\ }{{\bf #1} {(#2)} {#3}}}
\nc{\pl}[3]{{\it  Phys.\ Lett.\ }{{\bf #1} {(#2)} {#3}}}
\nc{\prep}[3]{{\it Phys.\ Rep.\ }{{\bf #1} {(#2)} {#3}}}
\nc{\prsl}[3]{{\it Proc.\ R.\ Soc.\ London\ }{{\bf #1} {(#2)} {#3}}}
\nc{\ptp}[3]{{\it  Prog.\ Theor.\ Phys.\ }{{\bf #1} {(#2)} {#3}}}
\nc{\ptps}[3]{{\it  Prog\ Theor.\ Phys.\ suppl.\ }{{\bf #1} {(#2)} {#3}}}
\nc{\physa}[3]{{\it  Physica\ A\ }{{\bf #1} {(#2)} {#3}}}
\nc{\physb}[3]{{\it  Physica\ B\ }{{\bf #1} {(#2)} {#3}}}
\nc{\phys}[3]{{\it Physica\ }{{\bf #1} {(#2)} {#3}}}
\nc{\rmp}[3]{{\it  Rev.\ Mod.\ Phys.\ }{{\bf #1} {(#2)} {#3}}}
\nc{\rpp}[3]{{\it Rep.\ Prog.\ Phys.\ }{{\bf #1} {(#2)} {#3}}}
\nc{\sjnp}[3]{{\it Sov.\ J.\ Nucl.\ Phys.\ }{{\bf #1} {(#2)} {#3}}}
\nc{\spjetp}[3]{{\it Sov.\ Phys.\ JETP\ }{{\bf #1} {(#2)} {#3}}}
\nc{\yf}[3]{{\it Yad.\ Fiz.\ }{{\bf #1} {(#2)} {#3}}}
\nc{\zetp}[3]{{\it Zh.\ Eksp.\ Teor.\ Fiz.\  }{{\bf #1}  {(#2)} {#3}}}
\nc{\zp}[3]{{\it Z.\ Phys.\ }{{\bf #1} {(#2)} {#3}}}
\nc{\ibid}[3]{{\sl ibid.\ }{{\bf #1} {#2} {#3}}}
\nc{\rf}[1]{(\ref{#1})}
\nc{\nn}{\nonumber \\*}
\nc{\bfB}{\bf{B}}
\nc{\bfv}{\bf{v}}
\nc{\bfx}{\bf{x}}
\nc{\bfy}{\bf{y}}
\nc{\vx}{\vec{x}}
\nc{\vy}{\vec{y}}
\nc{\oB}{\overline{B}}
\nc{\oI}{\overline{I}}
\nc{\oR}{\overline{R}}
\nc{\rar}{\rightarrow}
\nc{\ti}{\times}
\nc{\slsh}{\hskip-5pt/}
\nc{\sm}{Standard~Model~}
\nc{\MP}{M_{\rm Pl}}
\nc{\tp}{t_{\rm Pl}}
\nc{\ave}{\bar{E}}

\nc{\eff}{{\rm eff}}
\nc{\kk}{\vek{k}}
\nc{\pp}{{\rm p}}
\nc{\ga}{g_{a\gamma}}
\nc{\vv}{\\}
\nc{\eee}{{\bf E}}
\nc{\bbb}{{\bf B}}
\nc{\qcd}{T_{\rm QCD}}
\nc{\G}{\rm \ G}
\def\vec#1{{\bf #1}}

\def\lae{\;^{<}_{\sim} \;} \def\gae{\; ^{>}_{\sim} \;} 

\def\ell{e^{c}LL}

\begin{document}
{\title{\vskip-2truecm{\hfill {{\small \\
	\hfill \\
	}}\vskip 1truecm}
{Conditions for a Successful Right-Handed Majorana Sneutrino Curvaton}}
{\author{
{\sc John McDonald$^{1}$}\\
{\sl\small Dept. of Mathematical Sciences, University of Liverpool,
Liverpool L69 3BX, England}
}
\maketitle
\begin{abstract}
\noindent

               We consider the conditions which must be satisfied for a 
Majorana RH sneutrino, a 
massive right-handed (RH) sneutrino associated with the see-saw 
mechanism of Majorana neutrino masses,  
to play the role of 
the curvaton. Planck-scale suppressed non-renormalizable terms in the 
RH neutrino superpotential must be eliminated to a high order
if the RH sneutrino curvaton is to dominate the energy density before it decays, 
which can be achieved via an R-symmetry which is broken to R-parity by the
 Giudice-Maseiro mechanism. 
In order to evade thermalization of the curvaton condensate, one RH neutrino mass 
eigenstate must have small Yukawa couplings, corresponding to a lightest
 neutrino mass $m_{\nu_{1}} \lae 10^{-3} \eV$. 
A time-dependent lepton asymmetry will be induced in the RH sneutrino 
condensate by the Affleck-Dine mechanism driven by SUSY breaking B-terms 
associated with the RH neutrino masses. 
Requiring that the 
resulting baryon asymmetry and isocurvature perturbations
 are acceptably small imposes an upper bound on the 
RH neutrino mass. We show that a scenario consistent with all constraints
 is obtained for a RH neutrino mass in the 
range $10^{2} - 10^{4} \GeV$ when the RH sneutrino decay temperature
 is greater than
 $100 \GeV$ and lightest neutrino mass 
$m_{\nu_{1}} \approx 10^{-3} \eV$. Larger RH neutrino masses
 are possible for smaller $m_{\nu_{1}}$. The resulting scenario is 
 generally consistent with a solution of the cosmic string problem of 
D-term inflation.

\end{abstract}
\vfil
\footnoterule
{\small $^1$mcdonald@amtp.liv.ac.uk}

\thispagestyle{empty}
\newpage
\setcounter{page}{1}

\section{Introduction}

                 The possibility that the density perturbations
 responsible for structure formation 
come from quantum fluctuations of a scalar field other than the 
inflaton has recently attracted considerable interest \cite{curv,moroi,curv2,clist}.
 The scalar field, 
known as the curvaton \cite{curv2}, is effectively massless
 during inflation, but begins coherent oscillations after inflation
 and comes to dominate the
 energy density of the Universe before it decays. The initial 
isocurvature perturbations of the 
curvaton are then converted to adiabatic
 energy density perturbations. A number of curvaton
 candidates have been proposed and analyses of curvaton dynamics 
performed \cite{clist,gordon,postma,snu1,enqvist,chaoticj,phaz1}. 

      An important question is 
whether there is a natural curvaton candidate in the context of the 
minimal supersymmetric (SUSY) Standard Model (MSSM) and its extensions. 
SUSY is advantageous as it allows us to understand the 
 flatness of the curvaton potential with respect to quantum corrections.
A specific motivaton for a SUSY curvaton is provided by 
 D-term hybrid inflation \cite{dti}. 
In D-term inflation with naturally large values of the superpotential coupling, 
the energy density during inflation must be lowered in 
order to 
evade the problem of cosmic microwave background (CMB)
 perturbations due to cosmic strings \cite{dtstring}.
  This in turn requires an 
alternative source of the cosmological energy density perturbations, such as a curvaton. 

          Some general conditions for a SUSY flat 
direction scalar to play the role of a curvaton were considered in \cite{postma}. 
Similar conditions for the 
specific case of right-handed (RH) sneutrinos (the scalar superpartners
 of the RH neutrinos) were discussed in 
\cite{snu1}. It was shown that small couplings of the curvaton  
to the MSSM fields are required in order to evade thermalization
 of the curvaton condensate by scattering from
 particles in the thermal background and in order to have 
a low enough decay temperature for the curvaton condensate 
to dominate the energy density 
before it decays. Curvaton domination also requires the elimination  
of Planck-scale suppressed non-renormalizable terms (NRT) to a high order 
in the flat direction superpotential. 

                  The case of a coherently oscillating MSSM flat direction 
condensate formed from a linear combinations of squark, slepton and Higgs fields 
was considered in \cite{enqvist}. It was shown that as a result of their interaction with 
the fields in the background radiation, 
 the condensates are either thermalized 
or fail to dominate the energy density before they decay. 
MSSM flat direction scalars can therefore play the role of a 
curvaton only if one assumes that the 
inflaton can primarily decay to 'hidden radiation' consisting of hypothetical 
light hidden sector fields \cite{enqvist}.  In the
following we will consider the conventional case of inflaton
decay to observable radiation. 

     An alternative curvaton candidate, in the context of the MSSM extended to 
accomodate neutrino masses, is a RH sneutrino. The curvaton
condensate is most likely to survive in the thermal enviroment
of the early Universe and to dominate the energy density before 
it decays if it is relatively weakly coupled to MSSM fields. This makes a 
RH sneutrino a promising candidate for a 
SUSY curvaton. This possibility was noted in 
\cite{postma} and discussed in \cite{moroi,snlept},
 which also considered the possibility of leptogenesis
 via RH sneutrino decay and the associated 
isocurvature fluctuations. In \cite{snu1} the  
conditions for a RH sneutrino to serve as a curvaton
were considered in some detail. 

            The results of \cite{snu1} showed that a
 Dirac RH sneutrino (a RH sneutrino associated with Dirac 
neutrino masses, corresponding to RH neutrino masses $M_{N} = 0$), 
could serve as a curvaton if the amplitude of the RH sneutrino at the 
onset of its coherent oscillations satisfied $N_{osc} \gae 
10^{-5} M$, where $M = M_{Pl}/\sqrt{8 \pi}$ with $M_{Pl}$ the Planck mass.
 This requires that  
Planck-scale suppressed superpotential NRT $\propto N^{n}$
 are suppressed for $n < 6$ \cite{snu1}. 
The weak Yukawa coupling of the Dirac RH sneutrino 
ensures that there is no danger of condensate thermalization. However, the 
decay temperature of the RH sneutrino condensate was found to be around 10 MeV, 
which introduces two problems \cite{snu1}. The decay of the condensate 
RH sneutrinos 
will produce a large number of lightest SUSY particles (LSPs),
 which will overclose the Universe unless 
they decay sufficiently rapidly. In order that their 
decay products do not destroy the light elements produced in 
nucleosynthesis, the RH sneutrinos and LSPs must decay
 before nucleosynthesis occurs at $T_{nuc} \approx 1 \MeV$. 
This requires new R-parity violating interactions in order to have 
sufficiently 
rapid LSP decay. (This also has the disadvantage of eliminating the 
possibility of LSP dark matter.) 
The second problem is that the baryon asymmetry 
must be produced during the decay or after the decay of the condensate
RH sneutrinos, since any 
pre-existing baryon asymmetry will not be correlated with 
the entropy fluctuations produced by curvaton decay, resulting in
 unacceptably large baryon isocurvature fluctuations \cite{gordon}. 
Therefore new baryon number violating interactions must be introduced 
in order that the 
RH sneutrino decay can directly produce the baryon asymmetry. 
Whilst no definitive analysis of these possibilities has yet been undertaken, 
the Dirac RH sneutrino curvaton appears to be disfavoured.           

           Majorana RH sneutrinos (RH sneutrinos associated with 
the see-saw mechanism for Majorana neutrino
 masses \cite{seesaw}, corresponding to the case $M_{N} \neq 0$) can also dominate the 
energy density of the Universe before they decay. However, because of 
the larger Yukawa couplings in the see-saw mechanism for neutrino masses, 
the RH sneutrino condensate decays earlier, requiring a larger initial 
amplitude, $N_{osc} \gae 10^{-3} M$, and correspondingly
 greater suppression of the 
Planck-suppressed superpotential 
NRT (all superpotential terms $\propto N^{n}$ with $n < 11$ \cite{snu1}).
 In addition, the Majorana RH sneutrino 
curvaton was shown to have a potential problem with condensate 
thermalization.  
Due to its larger Yukawa couplings, the interaction of the 
Majorana RH sneutrino curvaton condensate with the MSSM fields in the thermal 
background is much stronger than in the case of the Dirac RH sneutrino. 
As a result,  
any contribution to the effective light 
neutrino mass matrix resulting from couplings to the 
curvaton's fermionic partner must be less than $10^{-3} 
\eV$ in order to avoid thermalization of the condensate \cite{snu1}.
Thus if the masses of the neutrinos are
 characterized by the mass squared splittings suggested by solar and 
atmospheric neutrino oscillations, $m_{\nu} \sim 0.01 - 0.1 \eV$ 
\cite{moha}, the RH sneutrino condensate would be thermalized before it could
dominate the energy density. This will be the case if there is 
no correlation between the RH neutrino mass matrix and the 
neutrino Yukawa coupling matrix, in the sense that diagonalization of the
 RH neutrino mass matrix does not result in approximate diagonalization of the
 neutrino Yukawa coupling matrix.
 In this case the RH sneutrino curvaton mass 
eigenstate would have Yukawa couplings to the Higgs and lepton doublets 
characterized by the largest neutrino mass eigenvalue, $m_{\nu} \approx 
0.1 \eV$, ruling out the Majorana RH sneutrino as a curvaton.

    However, it is possible for mass squared splittings and mixing angles 
associated with solar and atmosperic neutrino
phenomenology to be accounted for by just two RH neutrino mass
 eigenstates. For example, this is the case in the single right-handed neutrino dominance
 (SRHND) model of neutrino masses \cite{king}. 
A third right-handed neutrino and its superpartner could 
 then have arbitrarily small Yukawa couplings 
to the MSSM fields, allowing its condensate to evade thermalization. In this case the 
Majorana RH sneutrino becomes a viable curvaton candidate. 
The requirement that the lightest neutrino mass satisfies 
$m_{\nu_{1}} \lae 10^{-3} \eV$ may then be regarded as a 
prediction of the Majorana RH sneutrino curvaton model.

              In this paper we consider in detail the case of a Majorana RH 
sneutrino curvaton with a light neutrino mass
eigenstate, which was not considered in \cite{snu1}. 
An advantage of the Majorana RH sneutrino
curvaton with respect to dangerous Planck-suppressed NRT 
is that the conventional R-symmetry of the MSSM (broken to R-parity by the 
$\mu H_{u} H_{d}$ term) \cite{nilles}, extended to include the RH neutrinos,  
 will completely eliminate all RH neutrino superpotential 
terms except the mass term and Yukawa coupling terms responsible for neutrino 
masses, thus solving the problem of dangerous Planck-suppressed NRT. The R-symmetry
 must be broken to R-parity, but if this 
occurs via the Giudice-Masiero mechanism \cite{gm}, which generates only
R-symmetry breaking effective superpotential terms with total R-charge equal to zero, 
then no dangerous RH neutrino superpotential NRT will arise. 
Operators eliminated by a global symmetry such as R-symmetry might be 
generated by quantum gravity wormhole effects \cite{worm}.
 However, this is dependent upon the short-distance structure
 of gravity \cite{lindworm}. 
In the following we will take the view that quantum gravity
 corrections are absent or highly suppressed. 

         We will show that a time-dependent 
lepton asymmetry is induced in the RH sneutrino 
condensate via a variant of the Affleck-Dine (AD) mechanism 
driven by the SUSY breaking B-terms associated with the RH neutrino masses.
(This mechanism has also recently been discussed in the more general context
 of AD leptogenesis via a RH sneutrino condensate in \cite{aldrees}.)
 Unlike the conventional AD mechanism associated with MSSM 
flat direction fields \cite{ad}, L-violation from the B-terms
remains unsuppressed throughout the evolution of the condensate, resulting in
 a rapidly oscillating lepton asymmetry. 
If the RH sneutrino decay occurs before the
 electroweak phase transition, as is likely to be necessary in order to generate
 the baryon asymmetry, then any lepton asymmetry will be
 processed into a baryon asymmetry 
of the same order of magnitude by sphalerons, which
 must be less than the observed baryon asymmetry of the Universe. 
The requirement that the 
lepton number to entropy ratio due to the decay of the 
RH sneutrino condensate satisfies $n_{L}/s \lae 10^{-10}$
 will be shown to impose a significant constraint on the Majorana RH sneutrino
 curvaton scenario. In particular, it will be shown that RH sneutrino
 leptogenesis via CP violating decays of the 
condensate RH sneutrinos \cite{snlept}
 can account for the baryon asymmetry only if the lightest 
neutrino mass is less than about $10^{-8} \eV$. This favours
 electroweak baryogenesis \cite{ewb} as the origin of the baryon asymmetry 
in this scenario.

            The paper is organized as follows. In Section 2 we discuss the conditions 
for a Majorana RH sneutrino to dominate the energy density and to 
account for the observed energy density perturbations. We also discuss the possibility 
that a Majorana RH sneutrino could account for the energy density perturbations in
 D-term inflation.  
In Section 3 we discuss the constraints on the RH neutrino Yukawa
couplings and neutrino masses coming 
 from evasion of condensate thermalization. 
We also discuss the possibility that thermal effective masses could play a role in the 
evolution of the condensate. In Section 4 we 
consider the AD lepton asymmetry induced in the Majorana RH sneutrino 
condensate by the B-term AD mechanism and the associated constraints on the
 model. In Section 5 we discuss the range of model parameters 
allowed by the constraints. In Section 6 we present our conclusions. 

\section{Curvaton Dominance and Density Perturbation Constraints} 

            We consider the simplest inflation
 scenario, in which the expansion rate during inflation, 
$H_{I}$, is constant and inflation is followed by a period of coherent
 inflaton oscillations.  
(We refer to this period as inflaton matter domination, IMD). Inflaton 
oscillations decay away completely at the reheating
 temperature, $T_{R}$, corresponding to 
the temperature at which the Universe becomes radiation dominated.  

           We will consider throughout the case where the reheating temperature
 from inflaton decay satisfies the thermal gravitino
 constraint \cite{grav}, $T_{R} \lae 10^{8} \GeV$  for $m_{3/2} = 100 \GeV$ (
$T_{R} \lae 10^{9} \GeV$  for $m_{3/2} = 1 \TeV$)
 \cite{grav2}; in the following we will consider the constraint
 for $m_{3/2} = 100 \GeV$. 
 This constraint is not necessary if the curvaton dominates the
 energy density for long enough
 that it dilutes the thermal gravitinos due to inflaton
 reheating. However, we will be
 considering the limiting case where the curvaton just dominates the energy density
 before it decays, so we will impose the thermal gravitino constraint for consistency. 
In this case the expansion rate at the end of IMD
will satisfy $H(T_{R}) \lae 10^{-2} \GeV$ $(1 \GeV)$
 for $T_{R} \lae 10^{8} \GeV$ ($10^{9} \GeV$),
 which is small compared with the value
 $H_{osc} \approx m_{N} \gae m_{s} \approx 100 \GeV$ 
 at which RH sneutrino coherent oscillations begin. Here
 $m_{N}$ is the RH sneutrino mass, $m_{s}$ is the gravity-mediated soft SUSY
 breaking scalar mass and $H(T_{R}) = k_{T} T_{R}^{2}/M$ is the expansion rate at
 the onset of radiation 
domination, where $k_{T} = (\pi^{2} g(T)/90)^{1/2}$ and 
$g(T)$ is the number of degrees of freedom in thermal equilibrum at $T$ \cite{eu}. 
 (In the following we will consider $g(T) \approx 200$, corresponding to all 
MSSM fields in equilibrium, such that $k_{T} \approx 4.7$.) 
Thus we will consider the RH sneutrino oscillations to begin during IMD.    

        The superpotential of the mass eigenstate 
RH neutrinos, $N_{i}$, is given by 
\be{e1} W_{\nu} = \lambda_{\nu \;ij} N_{i} H_{u}L_{j}
 + \frac{M_{N_{i}} N_{i}^{2}}{2}   ~,\ee
where $H_{u}$ and $L_{i}$ are the MSSM Higgs and
 lepton doublets respectively and $i$ is a generation index.  
This leads to Majorana neutrino masses via the see-saw mechanism \cite{seesaw}.
In addition, we generally expect Planck-scale suppressed 
non-renormalizable
 superpotential terms involving the RH neutrino superfields of the form 
\be{e2} W = \frac{\lambda_{n} N_{i}^{n}}{n! M^{n-3}}               ~.\ee
In \cite{snu1} it was shown that such terms must be eliminated
 for all $n < 11$ if the initial amplitude of the Majorana RH sneutrino 
curvaton is to be large enough for the coherently 
oscillating curvaton to dominate the energy density 
before it decays. Otherwise the contribution of Planck-suppressed
 terms to the effective mass squared 
of the RH sneutrino 
will be greater than $H^{2}$, thus driving the RH sneutrino amplitude to
 small values. This elimination
of the Planck-suppressed NRT requires an explanation 
in terms of a symmetry or other mechanism. One possibility is an R-symmetry.
The superpotential \eq{e1} is R-symmetric under a simple extension of the 
R-symmetry of the MSSM under which $R(Q,L,u^{c},d^{c},e^{c}) = 1$ and 
$R(H_{u},H_{d}) = 0$ \cite{nilles}. By assigning $R(N) = 1$, all the terms in 
\eq{e1} are permitted by the R-symmetry whilst all superpotential terms
 in \eq{e2} with $n \neq 2$ are excluded. The 
$\mu H_{u} H_{d}$ term, essential for satisfactory electroweak
 symmetry breaking \cite{nilles}, 
breaks this R-symmetry to R-parity. The 
 Giudice-Masiero mechanism \cite{gm} initially eliminates the $\mu$-term
 via an R-symmetry. 
The $\mu$-term is then generated by R-symmetry breaking
 due to a non-minimal kinetic term in the K\"ahler potential 
 of the form $z^{\dagger}H_{u}H_{d}$, coupling the SUSY breaking 
hidden sector superfield $z$ to the Higgs doublets. Since $R(z) = 0$, this term is permitted 
by the R-symmetry. The F-term of the $z$ superfield breaks both SUSY and
the R-symmetry, resulting in a $\mu$-term of the order of the electroweak scale 
in the superpotential of the low energy effective theory \cite{gm}.
 The important feature of this
 mechanism is that it only generates R-symmetry breaking terms in the low-energy
effective superpotential which have total R-charge equal to zero. 
As a result no superpotential terms of the form $N^{n}$
 ($n \neq 2$) will arise. However, this 
mechanism does require that some 
interactions between the hidden sector and
 observable sector superfields are eliminated
by a mechanism other than R-symmetry,
 since R-symmetry permits a superpotential 
term of the form $zN^{2}$, which would typically lead to a very 
large RH neutrino mass due to the $z$ expectation value $\approx M$. 

       The scalar potential of the mass eigenstate RH sneutrinos after inflation is given by
\be{e3} V(N) = \left( \frac{m_{N}^{2} - c H^{2} }{2} \right) |N|^{2} + 
B \left( \frac{M_{N} N^{2}}{2}  + h.c. \right)    ~,\ee
where we have suppressed the generation indices.
Here $m_{N}^{2} = m_{s}^{2} + M_{N}^{2} + m_{eff}^{2}$, where 
$m_{eff}^{2} = \lambda_{\nu}^{2} <\phi^{2}>$
 is the effective mass squared term due to the interaction of the RH sneutrino with 
thermal background MSSM fields, $\phi$ \cite{ellistherm}. The latter will be shown 
 to play no role in the evolution of the RH sneutrino curvaton. 
Since there is little motivation
 to study RH neutrino masses less than the order of $100 \GeV$, for simplicity 
we will restrict attention to the case $M_{N} \gae m_{s}$
 in the following and therefore consider
 $m_{N} \approx M_{N}$ throughout. In fact, we will show
that the allowed values of $M_{N}$ are generally
 larger than O(100)$\GeV$ in models where the 
curvaton condensate decays before the electroweak phase transition, which
 are the most promising with respect to baryogenesis.

           The mass squared terms and possibly also B-terms
 gain corrections during IMD due to non-zero inflaton F-terms \cite{h2,h22,drt}. 
(During inflation such terms must be effectively zero in 
order to have a scale-invariant density perturbation spectrum, which
 is natural in D-term inflation \cite{dti}.) 
We will consider a negative order $H^{2}$ correction after inflation with 
$c \approx 1$. This is because the evolution of
 the RH sneutrino under the influence of a positive mass squared
 term in the case $c \approx -1$ will 
suppress the initial amplitude of the RH sneutrino oscillation relative to its amplitude 
at the end of inflation. 
As we show later, this would require a large value of $N$
 during inflation in order for the curvaton to dominate the
 energy density before it decays, which would in turn require a large value of
 $H_{I}$ in order 
to have sufficiently large curvaton energy density perturbations. 
However, this large value of $H_{I}$ will generally result in 
excessively large inflaton energy density perturbations and gravitational waves.  

                   The RH sneutrino oscillations begin at $H_{osc}
 \approx M_{N}/c^{1/2} \approx M_{N}$.
The energy density of the RH sneutrino condensate at $a >a_{R}$ is then \cite{snu1}
\be{e4} \frac{\rho_{N}}{\rho_{r}} =
 \left(\frac{a}{a_{R}}\right) \left(\frac{\rho_{N}}{\rho}\right)_{osc} 
= \left(\frac{a}{a_{R}}\right) \frac{N_{osc}^{2}}{6 M^{2}}    ~,\ee
where $a_{R}$ is the scale factor at the onset of radiation domination, 
$\rho_{r}$ is the radiation energy density, $ \rho \approx 3 H^{2} M^{2}$ 
is the energy 
density due to the coherently oscillating inflaton field during IMD and 
$\rho_{N} = M_{N}^{2}N_{osc}^{2}/2$ is the
 initial energy density in the RH sneutrino oscillations. 
RH sneutrino domination therefore occurs at \cite{snu1}
\be{e5} H_{dom} = \left(\frac{N_{osc}^{2}}{24 M^{2}}\right)^{2} H_{R}     ~,\ee
where $H \propto a^{-3/2}$ during IMD. In this expression we have defined curvaton 
domination to correspond to $\rho_{N} > 4 \rho_{r}$, in order to ensure 
complete domination of the energy density by the curvaton.

          We consider first the case of a single neutrino generation, which will later 
correspond to 
the RH sneutrino mass eigenstate with the weakest Yukawa couplings in \eq{e1}. The
 RH sneutrino decay rate is given by 
\be{e6} \Gamma_{d} \approx \frac{\lambda_{\nu}^{2}M_{N}}{4 \pi}  
= \frac{m_{\nu}M_{N}^{2}}{4 \pi v^{2}}     ~,\ee
where the neutrino mass is given by the see-saw expression
\be{e7}  m_{\nu} = \frac{\lambda_{\nu}^{2} v^{2}}{M_{N}}  ~\ee
and  $<H_{u}> = v/2$. 
The decay temperature of the RH sneutrino condensate is then 
\be{e8}   T_{d} \approx \left(\frac{\lambda_{\nu}^{2} 
M_{N} M}{4 \pi k_{T}}\right)^{1/2}   \equiv 
\left( \frac{m_{\nu} M_{N}^{2} M}{4 \pi k_{T} v^{2}} \right)^{1/2}    ~.\ee
Therefore
\be{e9a} M_{N} \approx 150 
\left( \frac{g\left(T_{d}\right)}{200}\right)^{1/4} 
\left( \frac{T_{d}}{100 \GeV}\right) 
\left( \frac{10^{-4}\eV}{m_{\nu}}\right)^{1/2} 
\left( \frac{v}{100 \GeV}\right) \GeV    ~.\ee
From \eq{e9a} we see that if
 $M_{N} \gae 100 \GeV$ and $m_{\nu}$ is not much smaller than 
$10^{-4} \eV$ then $T_{d} \gae 100 \GeV$.
 In this case there is 
a possibility of generating the baryon asymmetry
 either via electroweak baryogenesis \cite{ewb} or via leptogenesis
 by RH sneutrino decay \cite{snlept}. 

              The condition for the RH sneutrino condensate to
 dominate the energy density before it decays,
 $H_{dom} > \Gamma_{d}$, is given by
\be{e10} \frac{N_{osc}}{M} > 
2 \left( \frac{9}{\pi k_{T} }\right)^{1/4} m_{\nu}^{1/4} 
M^{1/4} \left(\frac{M_{N}}{T_{R} v} \right)^{1/2}      ~.\ee
Therefore 
\be{e11} \frac{N_{osc}}{M} > 4.0 \times 10^{-3} \beta \;\; ; \;\;\; \beta 
= 
\left( \frac{m_{\nu}}{10^{-4} \eV} \right)^{1/4}
 \left( \frac{10^{8} \GeV}{T_{R}} \right)^{1/2} 
\left(\frac{M_{N}}{v}\right)^{1/2}    
 ~.\ee 
 Thus if the mass of the lightest neutrino 
is not very much smaller than $10^{-4} \eV$ then $N_{osc}$
 cannot be very much smaller than $M$.

          A value of $N$ larger than $M$ implies that
 $\rho_{N} \approx -H^{2} N^{2} \approx -\rho$, 
leading to a breakdown of the assumption that the energy density is dominated by
inflaton oscillations after inflation. In addition, chaotic initial
 conditions with all initial gradient 
energy densities $ \lae M^{4}$ would imply that no field 
can have a value at the onset of inflation significantly larger than $M$ \cite{chaoticj}. 
We will therefore require that $N_{osc} \lae M$. From \eq{e11} a range of 
$N_{osc}$ compatible with curvaton domination then exists only if 
\be{e11a} M_{N} \lae 0.06 \left( \frac{10^{-4} \eV}{m_{\nu}}\right)^{1/2}
\left(\frac{v}{100 \GeV}\right)  T_{R}    ~.\ee 

            A second condition that the RH sneutrino curvaton 
must satisfy is that the curvaton energy density perturbations are of the 
correct magnitude to account for the CMB temperature fluctuations. 
The curvaton energy density perturbation is given by \cite{curv2}
\be{e12} \delta_{\rho} = \frac{2 \delta N}{N} 
\approx \left(\frac{H_{I}}{\pi N_{I}}\right)          ~,\ee
where $N_{I}$ is the amplitude of 
$N$ during inflation and $\delta N \approx H_{I}/2 \pi$
is the quantum fluctuation of the RH sneutrino. 
The observed perturbation is given by
\footnote{More precisely, the power spectrum of the 
curvaton fluctuation is ${\cal P}_{\delta N/N} \equiv  
H_{I}/2 \pi N_{I} = 3{\cal P}_{\zeta}^{1/2}/2$, where 
the observed value of curvature spectrum is  
${\cal P}_{\zeta}^{1/2} = 4.8 \times 10^{-5}$ \cite{curv2}.} 
$\delta_{\rho} \approx 1.4 \times 10^{-4}$ \cite{curv2}.
If $N$ does not evolve from the end of inflation then
$N_{I} \approx N_{osc} \gae 10^{-3} M$, 
which would imply that $H_{I} \approx \pi 
\delta_{\rho} N_{osc} \gae 10^{12} \GeV$. 
However, it was observed 
in \cite{snu1} that the RH sneutrino field will roll to larger values
 if the expected  negative order $H^{2}$ 
mass squared correction arises after
 the end of inflation. In this case $N_{osc} \gg N_{I}$ 
is possible, allowing smaller values of $H_{I}$ to be compatible with a RH sneutrino
curvaton. During IMD, $H \propto a^{-3/2}$. 
In this case the evolution of the RH sneutrino
 due to a $-c H^{2}$ mass squared term is
 given by $N \propto a^{\gamma}$, where \cite{snu1}
\be{e13} \gamma  = \frac{1}{2} \left[ - \frac{3}{2}
 + \sqrt{\frac{9}{4} + 4 c} \right]     ~.\ee
 Thus 
\be{e14}  \frac{N_{osc}}{N_{I}} = \left( \frac{a_{osc}}{a_{e}} \right)^{\gamma}    
= \left( \frac{H_{I}}{H_{osc}} \right)^{2 \gamma/3}     ~,\ee
where $a_{osc}$, $a_{e}$ are the scale factors at the onset of curvaton oscillations
and the end of inflation respectively. 
The amplitudes of $N$ and $\delta N$ due to a mass squared correction 
evolve in the same way, therefore $\delta N/N$ 
remains constant and \eq{e12} is unaltered \cite{snu1}. The 
curvaton domination condition \eq{e11} combined with \eq{e12} 
and \eq{e14} then implies that the lower bound on $H_{I}$ 
is given by
\be{e15}  H_{I}^{1 + 2 \gamma/3} \gae  4.0 \times 10^{-3}
\beta \pi \delta_{\rho} H_{osc}^{2 \gamma /3}  M    ~.\ee
The lower bound depends on $c$. If $c = 1$, for example, then $\gamma = 0.5$ and 
with $\delta_{\rho} = 1.4 \times 10^{-4}$
the lower bound on the expansion rate during inflation becomes
 \be{e16}  H_{I} \gae 9.3 \times 10^{9}    
\left( \frac{m_{\nu}}{10^{-4} \eV} \right)^{3/16}
 \left( \frac{10^{8} \GeV}{T_{R}} \right)^{3/8} 
\left(\frac{M_{N}}{100 \GeV}\right)^{5/8} 
\left(\frac{100 \GeV}{v}\right)^{3/8} 
   \GeV   ~.\ee 
Smaller values of $H_{I}$ are possible with larger values of $c$. For 
example, with $c=2.5$, corresponding to $\gamma = 1$, the lower bound becomes
 \be{e162}  H_{I} \gae 2.4 \times 10^{8}    
\left( \frac{m_{\nu}}{10^{-4} \eV} \right)^{3/20}
 \left( \frac{10^{8} \GeV}{T_{R}} \right)^{3/10} 
\left(\frac{M_{N}}{100 \GeV}\right)^{7/10} 
\left(\frac{100 \GeV}{v}\right)^{3/10} 
   \GeV   ~.\ee 

          For the case of a positive $H^{2}$
 correction to the RH sneutrino mass squared with $c \approx -1$, 
we see from \eq{e13} that $Re(\gamma) =  -3/4$ and so 
 $N_{osc}/N_{I} = (H_{osc}/H_{I})^{1/2}
\approx \left(M_{N}/H_{I}\right)^{1/2}$.
 With $N_{I}$ related to
 $H_{I}$ by \eq{e12}, $H_{I}$ is therefore given by  
\be{e163} H_{I} =  \frac{\pi^{2} \delta_{\rho}^{2} N_{osc}^{2}}{M_{N}}    ~.\ee
The lower bound on $N_{osc}$, \eq{e11}, then implies that
\be{e164} H_{I}  \gae 1.8 \times 10^{23} \left(\frac{m_{\nu}}{10^{-4} \eV}\right)^{1/2} 
\left(\frac{10^{8} \GeV}{T_{R}}\right)   ~.\ee
$H_{I}$ is therefore much larger than the upper bound from gravitational waves
 produced during inflation, $H_{I} \lae 10^{-5}M$ \cite{eu}.

                  One important application of a SUSY curvaton is to
D-term hybrid inflation \cite{dti}. It has been shown that the
contribution of $U(1)_{FI}$ strings to the CMB,
assuming naturally large values 
of the order of 1 for the superpotential coupling, 
is inconsistent with observations 
in the case where the energy density perturbations
 are explained by conventional inflaton perturbations, 
\cite{dtstring}. 
In order
 to evade this problem, the expansion rate during
 inflation must satisfy $H_{I} \lae 4 \times 10^{12} g \GeV$ \cite{dtstring}, 
where $g$ is the
 Fayet-Iliopoulos gauge coupling \cite{dti}. In this case
 a new source of energy density 
perturbations is required, since conventional inflaton energy density 
perturbations are too small.
 From \eq{e16},  requiring that $H_{I} \lae 4 \times 10^{12} g \GeV$ implies that
 a RH sneutrino curvaton with $c = 1$ can evade the D-term inflation 
cosmic string problem and 
 provide the energy density perturbations if 
\be{e16a} M_{N} \lae 1.6 \times 10^{6} g^{8/5}
 \left( \frac{10^{-4}\eV}{m_{\nu}} \right)^{3/10} 
 \left( \frac{T_{R}}{10^{8} \GeV} \right)^{3/5}
 \left( \frac{v}{100 \GeV} \right)^{3/5}
\GeV   ~, \ee
whilst for $c=2.5$,
\be{e16b} M_{N} \lae 1.0 \times 10^{8} g^{10/7} 
 \left( \frac{10^{-4}\eV}{m_{\nu}} \right)^{3/14} 
 \left( \frac{T_{R}}{10^{8} \GeV} \right)^{3/7}
 \left( \frac{v}{100 \GeV} \right)^{3/7}
\GeV    ~.\ee

\section{Thermal Constraints on Majorana Right-Handed Sneutrino Curvatons} 

           In this section we discuss the constraints coming from the 
requirement that the Majorana RH sneutrino condensate is not
 thermalized before the condensate 
decays. We also comment on the possibility that an effective RH sneutrino mass 
 due to the interaction of the curvaton RH sneutrino with the thermal background could play 
a role in the evolution of the condensate \cite{ellistherm}.

         In \cite{snu1} it was observed that a Majorana RH sneutrino condensate will be  
thermalized before it decays if the Yukawa couplings of the mass eigenstate
 RH neutrinos are all of the same order of magnitude.
 This would be expected if there was
 no correlation between the RH neutrino mass matrix and the 
Yukawa coupling matrix $\lambda_{\nu \;ij}$, in the sense that the 
basis in which the RH neutrino mass matrix is diagonalized is unrelated to 
the basis in which the Yukawa coupling matrix is diagonalized, 
since in this case all the mass eigenstate 
RH neutrino superfields would have similar Yukawa couplings to the lepton doublets. 
The thermalization rate would then be determined by the
 largest Yukawa coupling,    
corresponding to the neutrino mass indicated by atmospheric 
solar neutrino oscillations, $m_{\nu} \approx 0.1 \eV$. 

       However, it is possible for one of the RH neutrino mass
 eigenstates to have much weaker Yukawa couplings to the MSSM fields than
 those which are responsible for 
the mass splittings and mixing angles observed in solar and atmospheric neutrino
 oscillations \cite{king}. In this case there will be a 
 hierarchical light neutrino mass spectrum $m_{\nu_{1}} \ll m_{\nu_{2}} \ll m_{\nu_{3}}$, 
where $m_{\nu_{i}}$ ($i=1,2,3$) are the light Majorana neutrino mass eigenvalues \cite{moha}.
 It is then possible for coherent
 oscillations of the weakly coupled RH sneutrino 
to evade thermalization. Since coherent oscillations of the more strongly coupled RH sneutrinos 
will be thermalized by scattering from the thermal
 background MSSM fields, the
 weakly coupled RH sneutrino will naturally play the role of the curvaton. 

        For simplicity, we will consider the Yukawa couplings of the weakly coupled
 RH sneutrino mass eigenstate $N_{1}$ to lepton doublets 
$L_{i}$ to be all of the same order of magnitude,  
$\lambda_{\nu\;1i} \sim \lambda_{\nu}$. $\lambda_{\nu}$ will then 
determine the lightest neutrino mass $m_{\nu_{1}}$ via the 
see-saw mechanism with $N_{1}$. It will also determine 
the thermalization rate of the RH sneutrino condensate.
 We will drop the generation subscript and write $N_{1}$ as
 $N$ in the following.

        The condition for the RH sneutrino condensate to evade
 thermalization is that the scattering rate of the
 most strongly coupled thermal background MSSM fields from the
 zero-momentum condensate RH sneutrinos is less than the expansion
 rate $H$. The strongest 
scatterings will be from top quarks/squarks in the thermal
 background via Higgs exchange. The 
scattering cross-section is given by \cite{snu1} 
\be{e17} \sigma_{sc} \approx  \frac{\alpha_{\nu} \alpha_{t}}{E_{CM}^{2}}  
\approx \frac{\alpha_{\nu} \alpha_{t}}{M_{N} T}     ~,\ee
where $\alpha_{\nu,\;t} = \lambda_{\nu,\;t}^{2}/4\pi$, $\lambda_{t}$ is the top quark
Yukawa coupling and  
$E_{CM}^{2} = m_{N} E_{T}$ is the centre of mass energy of the
 scattering particles, where $E_{T} \approx T$ is the energy of the thermal background
 particle. The scattering rate from relativistic top quarks/squarks  is then 
$ \Gamma_{sc} = n \sigma_{sc} v  = n \sigma_{sc} $
where $n \approx g_{t}T^{3}/\pi^{2}$ is the number density of top quarks and squarks 
($g_{t} \approx 10$ is the number of degrees of freedom
 in top quarks and squarks including spin and 
colour) and $v = 1$ for relativistic thermal particles. 
Thus 
\be{e19}     \Gamma_{sc} = \frac{g_{t} \alpha_{\nu} \alpha_{t}}{\pi^{2}} 
\frac{T^{2}}{M_{N}}     ~.\ee 
The important feature of this expression is that the scattering
 rate is proportional to $T^{2}$, as is $H$ during radiation
 domination. 
Therefore the condition for the condensate to evade thermalization by scattering
 during radiation domination, $\Gamma_{sc} < H \equiv k_{T}T^{2}/M$, 
is independent of the temperature, 
\be{e20} \lambda_{\nu}^{2} \lae \frac{4 \pi^3 k_{T}}{g_{t} \alpha_{t}}
 \frac{M_{N}}{M}     ~.\ee
Using the relation between the Yukawa coupling $\lambda_{\nu}$ and the lightest 
neutrino mass $m_{\nu_{1}}$, \eq{e7}, then implies that
\be{e22} m_{\nu_{1}} \lae \frac{4 \pi^3 k_{T}}{g_{t} \alpha_{t}} 
\frac{v^{2}}{M} \approx  2 \times 10^{-3}
 \left(\frac{10}{g_{t}} \right) 
 \left(\frac{0.1}{\alpha_{t}} \right) 
\left(\frac{v}{100 \GeV} \right)^{2}  \eV
~,\ee 
where we have used  $k_{T} = 4.7$. 
From this we conclude that for $m_{\nu_{1}} \lae 10^{-3} \eV$ 
 it should be possible for the Majorana 
RH sneutrino curvaton to evade thermalization during the radiation dominated 
era, $T \leq T_{R}$. 

             This condition is also sufficient to ensure that the curvaton is 
not thermalized by the background radiation coming from inflaton decays
 during the IMD era. To see this, note that the 
expansion rate is related to the background radiation temperature
 during IMD by $T = k_{r} \left(M_{Pl} H T_{R}^{2} \right)^{1/4}$
 where $k_{r} = (9/5\pi^{3} g(T))^{1/8}$ \cite{eu}.
 Thus $H \propto T^{4}$ whilst $\Gamma_{sc}$ is proportional
 to $T^{2}$ as before. Therefore as $H$ and $T$ increase,
 the scattering rate
 becomes small relative to the expansion rate. 
Thus the largest scattering rate relative to the expansion rate
 will occur at the lowest temperature during
 IMD, $T_{R}$. However,
we know that $\Gamma_{sc} < H$ at $T_{R}$ if \eq{e22}
 is satisfied. Therefore 
if \eq{e22} is satisfied, the RH sneutrino condensate will evade
 thermalization during both the radiation dominated and IMD era. 

            A second thermal effect is the effective mass term generated 
by the interaction of the RH sneutrinos with
 the thermal background MSSM fields \cite{ellistherm}. This 
gives rise to a mass squared term of the
 form $\lambda_{\nu}^{2} <\phi^{2}>  \approx  
 \lambda_{\nu}^{2} T^{2}$, where $\phi$ is a $H_{u}$ or $L_{i}$ field in the thermal
 background and $<...>$ denotes the thermal average. Requiring that this is small compared 
with $M_{N}^{2}$ implies that $\lambda_{\nu} T_{max} < M_{N}$, where  
$T_{max}$ is the largest background temperature experienced by the 
coherently oscillating RH sneutrino. 
This corresponds to the temperature at the onset of 
RH sneutrino oscillations, $T_{max} = 
k_{r} \left(M_{Pl} H_{osc} T_{R}^{2}\right)^{1/4} \approx 
k_{r} \left(M_{Pl} m_{N} T_{R}^{2}\right)^{1/4}$, 
where $k_{r} \approx 0.4$ for $g(T) = 200$. 
Therefore 
\be{e25} \lambda_{\nu}  < \frac{M_{N}^{3/4}}{k_{r} T_{R}^{1/2} M_{Pl}^{1/4} } = 
1.4 \times 10^{-7} 
\left( \frac{ M_{N} }{100 \GeV} \right)^{3/4} 
\left( \frac{ 10^{8} \GeV }{T_{R}} \right)^{1/2}   ~.\ee
This translates into an upper bound on the lightest neutrino mass 
\be{e26} m_{\nu_{1}} \lae 2 \times 10^{-3} 
\left(\frac{v}{100 \GeV} \right)^{2}
\left(\frac{M_{N}}{100 \GeV} \right)^{1/2}
\left(\frac{10^{8} \GeV}{T_{R}} \right) \eV    ~.\ee 

         Thus $m_{\nu_{1}} \lae 10^{-3} \eV$ both allows the  
Majorana RH sneutrino condensate to evade thermalization
 and ensures that thermal effective masses
will play no role in the evolution of the RH sneutrino curvaton. The existence of a very 
light neutrino eigenstate may be regarded as a prediction of the
 Majorana RH sneutrino curvaton model.

\section{Lepton Asymmetry Constraints} 

    If the RH sneutrino condensate decays at $T_{d} \gae T_{ew}$ then we must ensure 
that any lepton asymmetry produced by the decay of the condensate, which would be
 converted into a baryon asymmetry of equal order of magnitude by sphaleron processes 
($n_{B}/s = -n_{L\;initial}/2s$ \cite{eu}), is not larger than the observed baryon 
asymmetry of the Universe. 

             In general, a coherently oscillating complex scalar field in SUSY
 will acquire a global charge asymmetry via the Affleck-Dine (AD) mechanism \cite{ad}. 
Soft SUSY breaking A- and B-terms in the scalar potential \cite{nilles} will cause the 
real and imaginary parts of the complex scalar field to oscillate out of phase, such that 
the scalar field descibes an ellipse in the complex plane corresponding to a globally 
charged scalar field condensate \cite{ad,phaz1}.
However, it is not possible to generate the baryon asymmetry 
 via AD baryogenesis or leptogenesis from the decay of a curvaton, 
since in this case there would be large baryon isocurvature
 perturbations due to the fact that in the AD mechanism both the real and
 imaginary parts of the curvaton field act to determine the baryon or lepton 
asymmetry \cite{adprob}. 

         The global $U(1)$ symmetry in the case of a RH sneutrino condensate
is lepton number L whilst the B-terms driving the Affleck-Dine
 mechanism are those associated with the RH neutrino 
mass terms in \eq{e1}. 
The B-term in \eq{e3} splits the masses of the real and
 imaginary parts of the RH sneutrino field. For $H^{2} \ll M_{N}^{2}$
 the scalar potential is given by  
\be{e28} V(N) = \left(M_{N}^{2}+ B M_{N}\right)\frac{N_{1}^{2}}{2} 
+ \left(M_{N}^{2}- B M_{N}\right)\frac{N_{2}^{2}}{2}   ~,\ee
where $N = (N_{1} + i N_{2})/\sqrt{2}$ and we have taken
 the RH sneutrino mass $m_{N}$ in \eq{e3}
to be equal to $M_{N}$. (We may choose the 
phase of $N$ such that $B$ is real and positive.) The solutions for coherent
 oscillations of the real and imaginary parts of the RH sneutrino field are
then $N_{1} = A_{1} \cos\left(m_{1}t\right)$ and $N_{2} = A_{2} 
\cos\left(m_{2}t + \delta\right)$,  
where $m_{1} = \left(M_{N}^{2} + B M_{N}\right)^{1/2}$, 
$m_{2} = \left(M_{N}^{2} - B M_{N}\right)^{1/2}$
and $A_{1,\;2} \propto a^{-3/2}$.
We have included a phase shift $\delta$, since this would be generally 
expected 
to arise during the initial transition from Hubble damped evolution to 
coherent oscillations. We will see that $\delta$ has a negligible  
effect on the net lepton asymmetry. 
Since the B-term in the scalar field equation for $N$ 
becomes dynamically significant, $BM_{N} \gae H^{2}$, 
only once $H \approx M_{N}$, corresponding to the 
onset of coherent oscillations, there is no reason to expect $N$
to be along the real direction 
at the onset of RH sneutrino oscillations. Thus we expect the phase of $N$ initially to be 
random and of the order of 1, such that $A_{1}$ and $A_{2}$ will be of the same order 
of magnitude. 

                 The lepton asymmetry density 
in the RH sneutrino condensate, $n_{L\;cond}$,  is given by
\be{e30} n_{L\;cond} = i \left(\dot{N}^{\dagger}N - N^{\dagger} \dot{N} \right) 
=  \left(\dot{N}_{2}N_{1} - \dot{N}_{1} N_{2}\right)    ~.\ee
Therefore
\be{e31} n_{L\;cond} = A_{1}A_{2} 
\left(m_{1}\sin(m_{1}t)\cos(m_{2}t + \delta)   
-  m_{2}\sin(m_{2}t + \delta)\cos(m_{1}t) 
\right)   ~.\ee
For example, for the case $\delta = 0$, if we consider $M_{N} \gg B$
then $m_{1} \approx M_{N} + 
\delta m/2$, 
$m_{2} \approx M_{N} - \delta m/2$, where $\delta m \approx B$. Then
\be{e32} n_{L\;cond}  \approx  A_{1}A_{2} m \left( \sin\left(\delta m  t\right) + 
\frac{\delta m}{2 m} \sin \left( 2 m t\right)   \right)    ~.\ee
Thus, for $m$, $\delta m \gg H(T_{d})$, there will be rapid oscillations
of the lepton asymmetry between large positive and negative lepton number
over the time scale of condensate decay, 
$\Gamma_{d}^{-1} \approx H^{-1}(T_{d})$.

         We next calculate the net lepton asymmetry
 produced by the decay of the RH sneutrino condensate. 
 Although the lepton asymmetry in the condensate is time dependent, once
 lepton number is transferred 
to the thermal background it will be {\it conserved}. 
(We neglect the effect 
sphaleron processing for now, which will not affect
 the magnitude of the lepton asymmetry.) 
This follows since lepton number violating 
scattering processes due to RH neutrino exchange 
are out of equilibrium at temperatures 
below $T_{f} \approx 1.2 \times 10^{11} (1 \eV/m_{\nu})^{2} \GeV$ \cite{cko}.  
Therefore the rate of increase of the thermal background 
lepton number density, $n_{L}$, from RH sneutrino condensate decay
is given by 
\be{e33}  \frac{1}{a^{3}} \frac{d \left(a^{3} n_{L}\right)}{d t}
 = \Gamma_{d} e^{-\Gamma_{d}t} n_{L\;cond}      ~,\ee
where we have included a factor $e^{-\Gamma_{d}t}$ in 
order to account for the decay of the RH sneutrino condensate and the 
scale factor terms account for the expansion of the Universe. 
Thus $$  
n_{L} = \Gamma_{d} A_{1}A_{2} 
\int_{0}^{\infty} e^{-\Gamma_{d}t} 
\left[
\left(
m_{1}\sin(m_{1}t)\cos(m_{2}t) 
 - m_{2}\sin(m_{2}t)\cos(m_{1}t)
\right) \cos \delta \right.  
$$
\be{e34}
\left.
-  
\left(
m_{1}\sin(m_{1}t)\sin(m_{2}t) 
 +
 m_{2}\cos(m_{1}t)\cos(m_{2}t)
\right) \sin \delta \right] dt   
~.\ee
The integrals may be calculated exactly, with the result
$$
n_{L} = 
\frac{\Gamma_{d}A_{1}A_{2}\left(m_{1}^{2}-m_{2}^{2}\right)
\left[\Gamma_{d}^{2} + m_{1}^{2} +m_{2}^{2}\right] \cos \delta }{
\left[ \Gamma_{d}^{2} + \left(m_{1}-m_{2}\right)^{2} \right] 
\left[ \Gamma_{d}^{2} + \left(m_{1}+m_{2}\right)^{2} \right] 
} 
$$
\be{e34a} 
- \frac{\Gamma_{d}^{2}A_{1}A_{2}
\left(m_{2} \Gamma_{d}^{2} + 3 m_{1}^{2}m_{2} 
+ m_{2}^{3} \right) \sin \delta}{
\left[ \Gamma_{d}^{2} + \left(m_{1}-m_{2}\right)^{2} \right] 
\left[ \Gamma_{d}^{2} + \left(m_{1}+m_{2}\right)^{2} \right] 
} 
~.\ee
For small $\Gamma_{d}^{2}$ compared with $\left(m_{1}-m_{2}\right)^{2}$,
and assuming that the natural value of $\delta$ is not very close to 
$\pi/2$, 
the term proportional to $\sin \delta$ is negligible and \eq{e34a}
becomes 
\be{e34b} n_{L} \approx \frac{ \Gamma_{d} 
A_{1} A_{2} \left(m_{1}^{2} + m_{2}^{2}\right) \cos 
\delta }{\left(m_{1}^{2} 
- m_{2}^{2}\right)} 
= \frac{\Gamma_{d}A_{1}A_{2}M_{N} \cos \delta}{B}  ~.\ee
We wish to express this in the form $n_{L}/s$. 
The energy density in the RH sneutrino condensate is given by 
\be{e34c} \rho = \rho_{1} + \rho_{2} = \frac{m_{1}^{2}A_{1}^{2}}{2} +
\frac{m_{2}^{2}A_{2}^{2}}{2}    ~.\ee
The entropy density at $T_{d}$ is related to $\rho$ by $s = 4 \rho/3T_{d}$ \cite{eu}, where 
we assume the radiation energy density from RH sneutrino decay at $T_{d}$ equals $\rho$.
Thus the lepton number to entropy density is given by 
\be{e36} \frac{n_{L}}{s} =  
\frac{3 T_{d} \Gamma_{d} M_{N} \cos \delta}{2 B}
 \frac{A_{1}A_{2}}{\left(m_{1}^{2}A_{1}^{2} + m_{2}^{2}A_{2}^{2}\right) } 
  ~.\ee
With $A_{1} \sim A_{2}$, $m_{1}^{2} \approx m_{2}^{2} \approx M_{N}^{2}$, 
$\cos \delta \approx 1$ 
  and $B \lae M_{N}$, this becomes
\be{e36a}   \frac{n_{L}}{s} \approx \frac{T_{d}\Gamma_{d}}{M_{N}B}    ~.\ee
Using $\Gamma_{d} = k_{T} T_{d}^{2}/M$, the lepton asymmetry from B-term 
AD leptogenesis is therefore given by
\be{e37} \frac{n_{L}}{s} \approx  
 \frac{k_{T} T_{d}^{3}}{ B M_{N} M}    ~.\ee
Requiring that $n_{L}/s \lae 10^{-10}$  imposes the constraint  
\be{e38}   T_{d} \lae 7.9 \times 10^{3} 
\left(\frac{B}{100 \GeV}\right)^{1/3}
\left(\frac{M_{N}}{100 \GeV}\right)^{1/3}
\GeV    ~.\ee
With $T_{d}$ given by \eq{e9}, this implies an upper bound on $M_{N}$, 
\be{e39} M_{N} \lae 1.3 \times 10^{5} 
\left(\frac{10^{-4} \eV}{ m_{\nu_{1}} }\right)^{3/4}
\left(\frac{B}{100 \GeV}\right)^{1/2}
\GeV   ~.\ee
This may also be expressed as an upper bound on $T_{d}$, 
\be{e39a} T_{d} \lae 8.6 \times 10^{4} 
\left(\frac{10^{-4} \eV}{ m_{\nu_{1}} }\right)^{1/4}
\left(\frac{B}{100 \GeV}\right)^{1/2}
\GeV  ~.\ee
Thus if $m_{\nu_{1}}$ is not very much smaller than the mass of the 
other neutrino eigenstates then the lepton asymmetry constraint \eq{e39} favours a
relatively light RH neutrino mass, $m_{N} \lae 10^{5} \GeV$, whilst \eq{e39a} implies that 
the decay temperature is relatively low, $T_{d} \lae 10^{5} \GeV$.  

        The analysis of the B-term AD mechanism given above is similar to that recently 
given in \cite{aldrees} in the more general context of AD
 leptogenesis from a RH sneutrino condensate.
 Some of the analytical results we have obtained here should be useful 
 in this more general context, in particular the expression for the lepton 
asymmetry resulting from
 decay of the condensate, \eq{e34a}. From \eq{e34a} we see that the 
maximum asymmetry will be obtained for
 $\Gamma_{d} \approx |m_{1} - m_{2}| \approx B$, 
in agreement with \cite{aldrees}. However, it is worth emphasizing 
that a significant asymmetry can be still obtained 
even if $\Gamma_{d} \ll B$, as is clear from \eq{e36a}. 

        There are two obvious possibilities for the 
origin of the baryon asymmetry in the Majorana RH sneutrino curvaton model:
 electroweak baryogenesis \cite{ewb} or 
RH sneutrino leptogenesis based on conventional 
leptogenesis via CP violating decays of the condensate RH sneutrinos \cite{snlept}. 
Both of these require that the curvaton condensate decays before the electroweak 
phase transition. In the case where the energy density is dominated by the 
RH sneutrino condensate, it was shown in \cite{snlept} that the lepton asymmetry
from CP-violating decays of the condensate RH sneutrinos is given by 
\be{e40} \frac{n_{L}}{s} \approx 0.7 \times 10^{-10}
 \left(\frac{T_{d}}{10^{6} \GeV}\right)
\left(\frac{m_{\nu_{3}}}{0.05 \eV}\right) \delta_{eff}   ~\ee
where $\delta_{eff} \lae 1$ is the CP-violating phase. 
Therefore RH sneutrino leptogenesis
requires that $T_{d} \approx 10^{6} GeV$ if $\delta_{eff} \approx 1$. From 
the lepton asymmetry constraint, \eq{e38}, this would require that 
$M_{N} \gae 10^{8} \GeV$. 
\eq{e39} in turn requires that 
$m_{\nu_{1}} \lae 10^{-8} \eV$. Therefore RH sneutrino leptogenesis is only possible 
if the lightest neutrino mass is extremely small compared with the mass scale 
suggested by solar and atmospheric neutrino oscillations. 

\section{Consequences of the Constraints} 

                  We next consider the consequences of the simultaneous application 
of the constraints derived above for the parameter space
 of the Majorana RH sneutrino curvaton model. 
 We will consider the allowed region of
 the $(M_{N},T_{R})$ parameter space for different   
values of $m_{\nu_{1}}$ satisfying the thermalization 
upper bound $m_{\nu_{1}} \lae 10^{-3} \eV$ and for various
 lower bounds on $T_{d}$.  

        The constraints we apply are: (i) the upper bound on $M_{N}$ from 
curvaton dominance combined with $N_{osc} \lae M$, \eq{e11a}, 
(ii) the lower bound on $M_{N}$ from requiring that
 the RH sneutrino decay temperature 
in \eq{e9a} is greater than a given value and (iii) the upper
 bound on $M_{N}$ from the AD leptogenesis 
constraint, \eq{e39}. In addition, we will consider the upper bound on 
$M_{N}$ for which a RH sneutrino curvaton  
is compatible with a solution of the D-term inflation
 cosmic string problem for the case $c = 1$, \eq{e16a}. 
(We consider $g = 1$ for the $U(1)_{FI}$ gauge coupling throughout.)

\begin{figure}[hp]
\begin{center}
\includegraphics[width=0.75\textwidth]{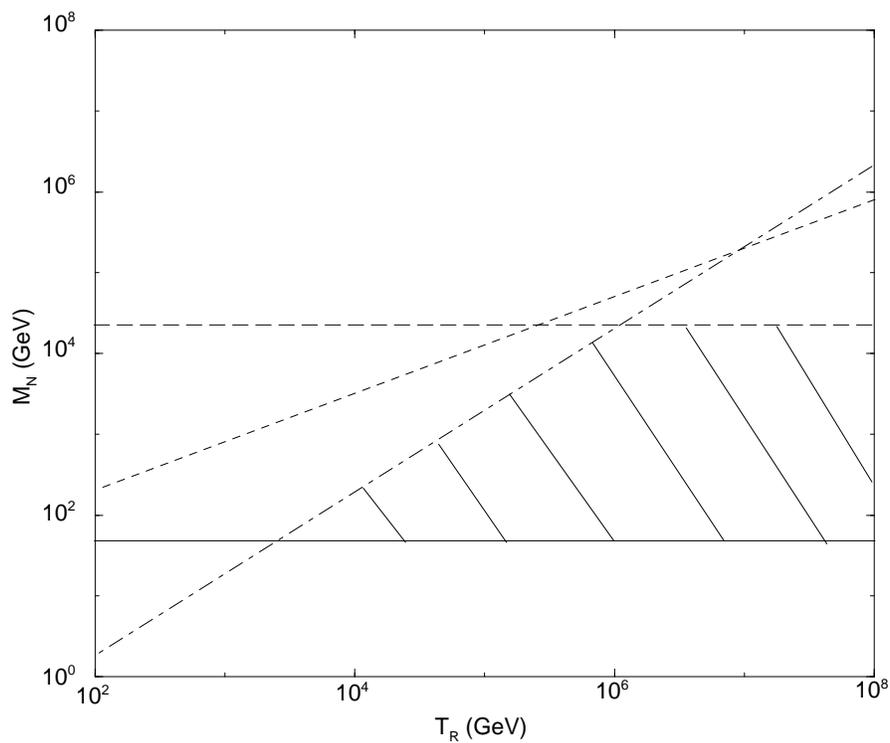}
\caption{\footnotesize{Right-handed neutrino mass vs. reheating 
temperature for $m_{\nu_{1}} = 10^{-3} \eV$ and $T_{d} > 10^{2} \GeV$. 
The solid line is the lower bound on $M_{N}$ from $T_{d} > 10^{2} 
\GeV$. The long-dashed line is the upper bound from B-term AD leptogenesis. The 
dot-dashed line is the upper bound from curvaton dominance. These 
define the allowed region, shown hatched. The short dashed line is the  
D-term inflation upper bound on $M_{N}$ for $c = 1$.  
}} \label{fig:} \end{center} \end{figure}

\begin{figure}[hp]
\begin{center}
\includegraphics[width=0.75\textwidth]{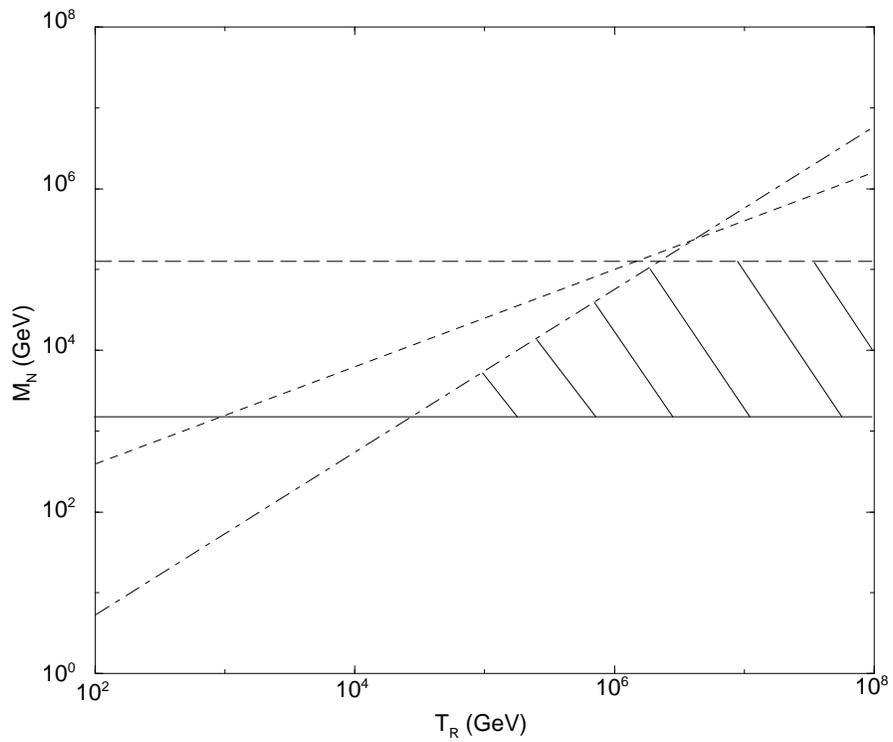}
\caption{\footnotesize{
Right-handed neutrino mass vs. reheating 
temperature for $m_{\nu_{1}} = 10^{-4} \eV$ and $T_{d} > 10^{3} 
\GeV$. 
}} \label{fig:} \end{center} \end{figure}

\begin{figure}[hp]
\begin{center}
\includegraphics[width=0.75\textwidth]{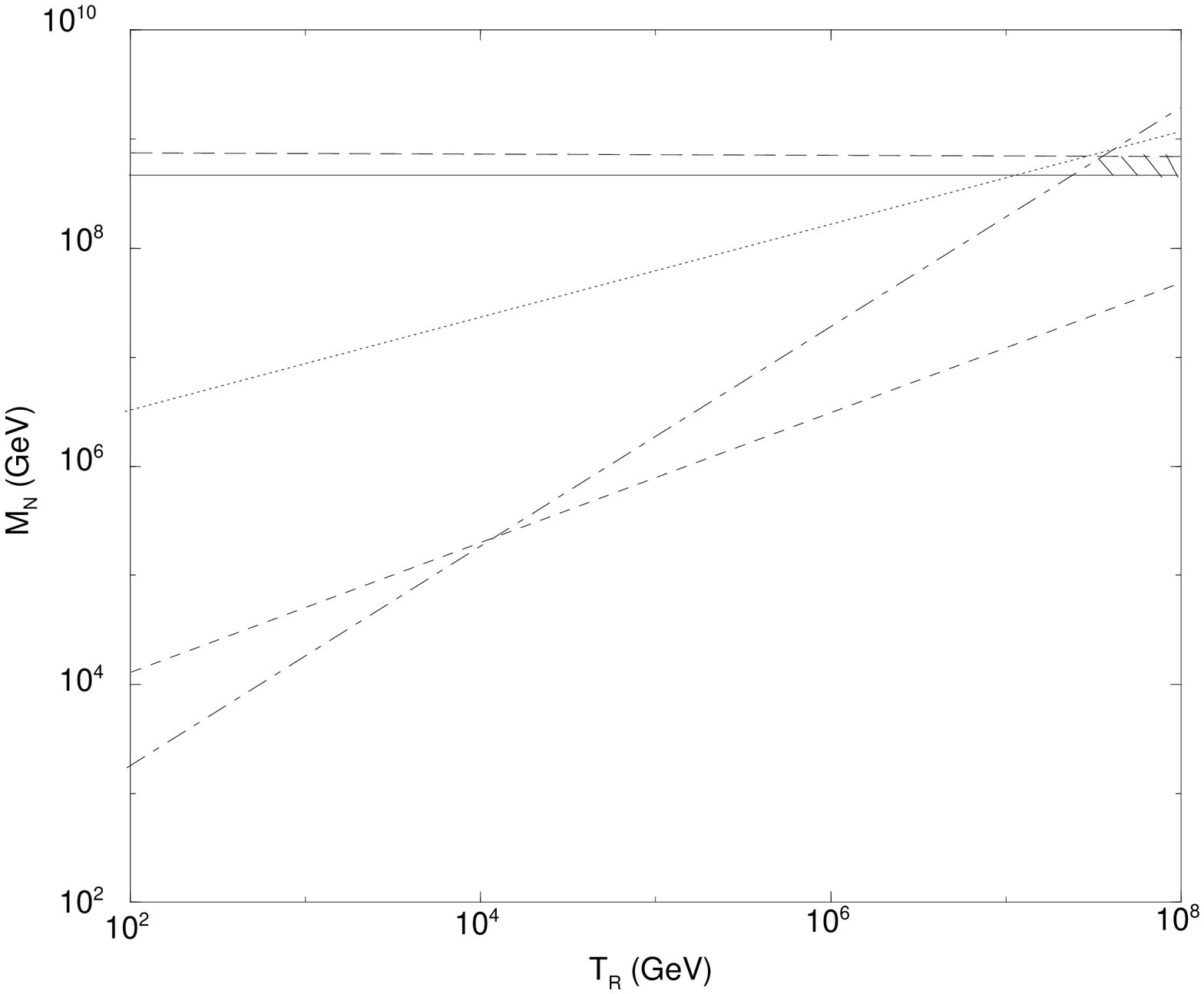}
\caption{\footnotesize{
Right-handed neutrino mass vs. reheating 
temperature for $m_{\nu_{1}} = 10^{-9} \eV$ and $T_{d} > 10^{6} 
\GeV$.  The D-term inflation upper bound for the 
case $c=2.5$ is also shown (dotted line). 
}} \label{fig:} \end{center} \end{figure}

      In Figure 1 we show the constraints 
for the case $m_{\nu_{1}} = 10^{-3} \eV$
(the largest value consistent with thermalization) 
and lower bound on $T_{d}$ equal to $100 \GeV$, corresponding to the 
value for which electroweak baryogenesis is possible.
The case $m_{\nu_{1}} \approx 10^{-3} \eV$ might be expected if 
$m_{\nu_{1}}$ is not very much smaller
 than $m_{\nu_{2,3}} \sim 0.01-0.1 \eV$. 
From Figure 1 the allowed range of $M_{N}$ is 
constrained to be between $47 \GeV$ and 
 $2.3 \times 10^{4} \GeV$. The allowed range of $M_{N}$ becomes narrower 
for $T_{R} < 10^{6} \GeV$, where the curvaton dominance
 constraint becomes more important than the AD leptogenesis constraint. 
 $T_{R}$ cannot be smaller than $2.7 \TeV$. 
It is interesting and perhaps significant 
that the allowed range of $M_{N}$ contains the $100 \GeV - 1 \TeV$ 
mass range, a natural mass range of particle physics. 
 
      In Figure 2 we consider a more restrictive possibility 
with $m_{\nu_{1}} = 10^{-4} \eV$ and $T_{d} > 1 \TeV$. The allowed range 
of $M_{N}$ is significantly narrower in this case compared with Figure 1, 
with $M_{N}$ 
between  $1.4 \times 10^{3} \GeV$ and $1.3 \times 10^{5} \GeV$. In general, increasing 
the lower bound on $T_{d}$ increases the lower bound on 
$M_{N}$ ($\propto T_{d}$, \eq{e9a}) whilst 
having no effect on the upper bound. Decreasing 
$m_{\nu_{1}}$ raises both the upper bound ($\propto m_{\nu_{1}}^{-3/4}$, 
\eq{e39}) 
and, to a lesser extent, the lower bound ($\propto m_{\nu_{1}}^{-1/2}$
\eq{e9a}).

      In Figure 3 we show the case where the lower bound on $T_{d}$ is 
$10^{6} \GeV$, compatible with RH sneutrino leptogenesis. 
In this case the lower bound on $M_{N}$ is  $5.0 \times 10^{8} \GeV$. 
We have shown the allowed parameter space for the case where 
$m_{\nu_{1}} = 10^{-9} \eV$, which is just compatible with 
RH sneutrino leptogenesis, as illustrated by the very small region of the 
parameter space compatible with the constraints. For 
$m_{\nu_{1}} \gae 10^{-8} \eV$ RH sneutrino leptogenesis 
is ruled out by the AD leptogenesis constraint. The case 
$T_{d} = 10^{6} \GeV$ is not 
compatible with a curvaton solution of the D-term inflation cosmic string 
problem when $c=1$. However, larger values of $c$ (but still of 
the order of 1) will allow a solution, as shown by the upper bound for 
the case $c = 2.5$.

         We conclude that unless $m_{\nu_{1}}$ is very small compared 
with the scale of neutrino masses responsible for solar and atmospheric 
neutrino oscillations, the 
Majorana RH sneutrino curvaton favours relatively 
low values of $M_{N}$ and $T_{d}$. 
The region of the parameter space compatible with a Majorana 
RH sneutrino curvaton is generally consistent with a curvaton solution 
of the D-term inflation cosmic string problem. 
RH sneutrino leptogenesis 
is ruled out as a source of the baryon asymmetry unless 
$m_{\nu_{1}} \lae 10^{-8} \eV$, thus favouring electroweak baryogenesis as 
the origin of the baryon asymmetry. However, it may be significant that 
RH sneutrino leptogenesis is not entirely ruled out as a source of the 
baryon asymmetry. 

\section{Conclusions}

              We have considered the conditions under which a Majorana RH sneutrino
could serve as a curvaton. One condition is the absence of Planck-scale suppressed 
corrections to the RH sneutrino scalar potential to a high-order. 
As the nature of such corrections requires 
speculation about the nature of the underlying full theory, we have
considered as one possibility the case where such corrections are absent. 
This might be explained by an R-symmetry, although this
requires an additional suppression of interaction terms between the  
observable sector and SUSY breaking hidden sector
 superfields in the superpotential.
A second condition is that the RH sneutrino 
condensate is not thermalized by scattering from the thermal background particles. 
This requires a model of neutrino masses in which one of the RH neutrino mass 
eigenstates has significantly weaker Yukawa couplings to the MSSM fields than the other 
RH neutrinos. 
The lightest neutrino mass eigenstate must then satisfy $m_{\nu_{1}}
 \lae 10^{-3} \eV$, which may be regarded as a prediction of the 
 scenario. The Majorana RH sneutrino curvaton scenario can therefore
 be ruled out experimentally 
by the observation of a degenerate neutrino mass spectrum,  $m_{\nu_{1}} \approx
m_{\nu_{2}} \approx m_{\nu_{3}}$.

           We have shown that a lepton asymmetry will be induced
 in the RH sneutrino condensate via the AD mechanism associated with SUSY 
breaking B-terms. Requiring that the baryon asymmetry coming from sphaleron 
conversion of this lepton asymmetry is smaller than the observed baryon asymmetry 
imposes an upper bound on $M_{N}$. 
Unless $m_{\nu_{1}}$ is very small compared with the mass scale
 of solar and atmospheric neutrino oscillations,  
relatively small values of $M_{N}$ ($ \lae 10^{5} \GeV$) are 
required for consistency with the AD leptogenesis constraint. 
 Some of the analytical results we have obtained here should be useful 
 in the more general analysis of AD leptogenesis from a
 RH sneutrino condensate.

     The AD leptogenesis constraint, combined with a lower bound on the 
RH sneutrino condensate decay temperature 
and the requirement that the RH sneutrino curvaton
condensate dominates the energy density before it decays, 
restricts the region of the $(M_{N}, T_{R})$ parameter space 
consistent with a RH sneutrino curvaton.
The range of  $M_{N}$ consistent with a RH sneutrino curvaton is 
typically smaller than that considered in the conventional see-saw mechanism where 
$\lambda_{\nu}$ is assumed to be of the order of the charged lepton
 Yukawa couplings, which requires $M_{N} \approx
 10^{9} \GeV$. However, the allowed range of $M_{N}$
for $m_{\nu_{1}} \approx 10^{-3} \eV$ and $T_{d} \gae 100 \GeV$
 is consistent with the $100 \GeV-1 \TeV$ mass scale, a natural mass scale
 in particle physics models.

              An important motivation for a SUSY curvaton is D-term hybrid inflation. 
 We have shown that the region of the parameter
space allowed by the constraints on the Majorana RH sneutrino curvaton 
is generally consistent with a solution of the D-term inflation cosmic 
string problem.             
D-term inflation naturally accounts for the absence of order $H^{2}$ corrections 
during inflation in both the inflaton and curvaton scalar potentials. Therefore a 
combined D-term inflation/Majorana RH sneutrino curvaton model could provide
a completely consistent model of SUSY inflation.
                         
                  The only obvious candidates for the origin of the baryon asymmetry 
in the Majorana RH sneutrino curvaton scenario are electroweak baryogenesis and 
RH sneutrino leptogenesis, AD leptogenesis being generally ruled out for a 
curvaton \cite{adprob}. We find that RH sneutrino leptogenesis via 
CP violating decays of the RH sneutrinos is consistent 
with the Majorana RH sneutrino curvaton only if the lightest neutrino mass eigenstate 
satisfies $m_{\nu_{1}} \lae 10^{-8} \eV$. Therefore RH sneutrino leptogenesis 
appears to be disfavoured, although it may be significant that it is not completely
 ruled out. This suggests that electroweak baryogenesis is the most likely source
 of the baryon asymmetry in this scenario.

\end{document}